# YouTube Video Analytics for Patient Engagement: Evidence from Colonoscopy Preparation Videos

Yawen GUO[a], Xiao LIU[b], Anjana SUSARLA[c], Rema PADMAN[d]
[a] University of California, Irvine
[b] Arizona State University
[c] Michigan State University
[d] Carnegie Mellon University

## Abstract

Videos can be an effective way to deliver contextualized, just-in-time medical information for patient education. However, video analysis, from topic identification and retrieval to extraction and analysis of medical information and understandability from a patient perspective are extremely challenging tasks. This study demonstrates a data analysis pipeline that utilizes methods to retrieve medical information from YouTube videos on preparing for a colonoscopy exam, a much maligned and disliked procedure that patients find challenging to get adequately prepared for. We first use the YouTube Data API to collect metadata of desired videos on select search keywords and use Google Video Intelligence API to analyze texts, frames and objects data. Then we annotate the YouTube video materials on medical information, video understandability and overall recommendation. We develop a bidirectional long short-term memory (BiLSTM) model to identify medical terms in videos and build three classifiers to group videos based on the levels of encoded medical information and video understandability, and whether the videos are recommended or not. Our study provides healthcare stakeholders with guidelines and a scalable approach for generating new educational video content to enhance management of a vast number of health conditions.

## 1 Introduction

The availability of medical information on the Internet, coupled with the advent of patients and clinicians alike using social media, has transformed how consumers access medical information to manage their illnesses. Medical information refers to information about physical, mental, or behavioral health or conditions created by healthcare providers or consumers in any form or medium, providing medical care to individuals or paying them (Fair Credit Reporting Act of 1992). Traditionally, patients receive such information and instructions in text format from clinicians. It has been proposed that purely text-based medical information can lead to reduced patient attention, understanding, recall, and compliance, especially for patients with low levels of literacy (Liu et al. 2014). Therefore, it is important to design patient education materials that increase patient engagement and participation in their medical care.

Videos are a valuable media format to educate patients and prepare caregivers in today's rapidly changing healthcare environment (ref). Patient education can facilitate knowledge acquisition, reduce anxiety, improve coping skills and enhance self-care behaviors (ref). Integrating visual and auditory information, videos are usually easy for individuals to understand and retain information (ref). For patients with chronic diseases who need complex medical information, videos are an effective means to deliver health advice and thereby improve the efficiency of care (Gagliano, M. E., 1988).

YouTube, the largest video sharing social media platform, provides user-generated medical information in a rich visual format, which may be easier to understand and comply with. YouTube hosts more than 100 million health related videos about the pathogenesis, diagnosis, treatment, and prevention of various medical conditions. Content on YouTube is created from both professional healthcare organizations as well as individuals (Madathil et al. 2015). For example, in the area of medical advice related to preparing for a colonoscopy exam, YouTube has videos ranging from colonoscopy prepping suggestions from medical institutions to videos from patients. The heterogeneity of information needs from patients and the ability to easily search and retrieve potentially misleading videos have exacerbated patient engagement challenges with medical information. Furthermore, they have eroded the value of YouTube as an educational channel for patients with chronic diseases, thereby making patients with limited health literacy less engaged and more vulnerable (Powers et al. 2017, Mackert et al. 2015).

This study establishes a YouTube video analysis pipeline involving educational interventions on patient health, using colonoscopy as an illustrative example. By creating and disseminating video-based patient educational materials for various conditions (including surgical, infectious and chronic conditions), YouTube may have great promise in changing the way healthcare works. From a macroeconomic perspective, the ability of patients to participate in medical information can not only improve resource utilization, but also improve the health of the entire population.

The paper is structured as follows. Section 2 describes the related prior research and background. Section 3 introduces the dataset and collection process. Sections 4 and 5 discuss the methods we use and the result analysis. Section 6 talks about the possible future work and limitations. Finally, Section 7 presents our conclusions.

## 2 Background

Patient education and empowerment is a topic of longstanding interest to healthcare practitioners as well as policy makers. In the past decade, policy makers have been optimistic about the role of social media in this regard (Hamm et al. 2013). Social media offers the public, patients, and health professionals a medium to communicate about health issues with the possibility of improving health outcomes (Moorhead et al. 2013). However, the challenge on sites such as YouTube is that anyone can publish a healthcare video, regardless of their background, medical qualification, professionalism, or intention. Therefore, health information available on YouTube can range from high quality videos to commercials or pseudo-scientific scams (Coiera et al. 2012; Lau et al. 2012; Syed-Abdul et al. 2013). Thus, there is a high probability of patients encountering misinformation during the information-seeking process (Hamm et al. 2013). There is also the danger that information available on YouTube could be incomplete or outdated, pushing patients into making potentially damaging decisions.

As the volume of online videos grows exponentially, using expert evaluation to assess the quality of all videos on YouTube is not a sustainable long-term solution. The frequency of views may be manipulated by parties with specific agendas to achieve "perceived" popularity. Videos may have higher view counts due to marketing campaigns, viral effects, because the video has been posted for a longer period of time or was linked from several web pages. Given the exponential growth of YouTube videos, an automatic multi-faceted approach that combines expert-driven (layperson, professional, and organizational-endorsement) and heuristic-driven criteria, could potentially be an ideal framework for assessing quality on YouTube.

Studies assessing the readability, suitability or comprehensibility of patient education materials on a myriad of topics abound, and the evidence is clear and consistent that most education materials are too complex for patients with low health literacy (ref). Patients' educational materials need to be understandable because many adults lack the requisite skills to obtain and process basic health information and services needed to make appropriate health decisions. The Patient Education Material Assessment Tool (PEMAT) assesses the domains of understandability. It is designed as a guideline to help determine whether patients will be able to understand and act on information. Patient education materials are understandable when consumers of diverse backgrounds and varying levels of health literacy can process and explain key messages (Shoemaker et al. 2013).

Colorectal cancer is the third most common cancer diagnosed in both men and women in the United States (ref). The percentage of the adult population that has been screened for colorectal cancer has increased over the last 10 years, largely because of an increase in colonoscopy procedures. The main benefit of getting a colonoscopy is that it helps detect early signs of cancer and allows your doctor to remove polyps which over time can become cancerous. During a colonoscopy, a patient goes under anesthesia and a doctor uses a probe to search the colon and rectum for precancerous polyps and cancer. Good bowel preparation for the procedure is essential for improved outcomes. Many patients are nervous about the discomfort or the possibility of pain because of the complexity of preparation and technical difficulties before and during the procedure. With proper educational interventions, patients can prepare well both physically and mentally and are able to self-manage the process. The success of patient education relies heavily on accessible medical information and patient-centered, personalized communication practices. Existing colonoscopy education is conducted in several ways and mainly aims to improve the quality of the procedure preparation. Prior studies show that interventions with cartoon visual aids, multimedia software, the telephone-based education, the physician-delivered education and a single educational video generated by a professional organization are efficacious to improve bowel preparation for colonoscopy. We focus on generating new colonoscopy educational materials from the corpus of YouTube videos and aim to give patients a comprehensive understanding of colonoscopy, not just the preparation procedure.

## 3 Data

### 3.1 Keyword Collection

To assess the medical information encoded in YouTube videos and their understandability, we develop a data collection process that first generates keywords about colonoscopy, and then retrieves the videos. We select the search keywords from three sources. The first is one of the largest online health communities, DailyStrength. In DailyStrength's Expert Answers forum, patients ask questions about colonoscopy, while certified medical experts provide answers. The search keywords used by patients on the Expert Answers forum represent typical informational needs of patients as shown in Figure 1. Terms are extracted from the posts and recommendations. The second source is a list of search suggestions appearing below the YouTube search bar with the search keyword 'colonoscopy'. Search on YouTube strives to surface the most relevant results according to keyword queries. Videos are ranked based on a variety of factors including how well the title, description, and video content match the viewer's query. For the third source, we search literature about colonoscopy education and collect keywords from papers. We take the union of these three sources and rank keywords by their popularity then select a set of keywords that cover the patient's most concerned topics.

Figure 1. Keyword Generation from Daily Strength

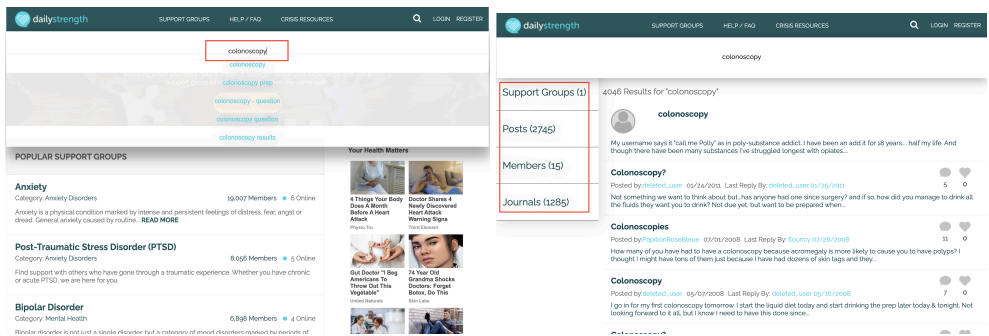

The project uses search keywords about colonoscopy and simulates patients' information seeking process as follows. With 20 keywords generated, 988 video IDs are collected, and 828 unique video IDs remain after removing duplicates because different keywords may lead to the same videos. We further filter videos by restricting the video language to be English. Some videos cannot be downloaded because of video license issues. Finally, we downloaded 312 videos out of 328 unique IDs and metadata are collected for downloaded videos.

## 3.2 Metadata Collection

We employ the YouTube Data API to retrieve video metadata. The information collected includes channel ID (account name), publishing time of the video, video title, video description, video tags, video duration, video definition, video caption availability, video rating, view count, like count, dislike count, and comment count. We select important information from video metadata and use them as features to build the final classifier for overall recommendation.

## 3.3 Annotation

The collected videos are annotated by three graduate research associates. They only have access to the video content but no other information on the YouTube webpage. All three annotated the videos according to the following criteria: medical information, understandability and overall recommendation.

*3.3.1 Medical Information*

*Does this video contain relevant medical information?* Medical information can be assessed along several critical dimensions, such as content understandability by end-users, the volume of medical information, the complexity of medical information provided, and so on. Medical information in online videos is conveyed through medical terminology, which constitutes healthcare-related words such as diseases, conditions, procedures, symptoms, and treatments. Raters annotate videos as encoding high or low levels of medical information from a patient perspective.

This criterion focuses on the volume of colonoscopy-related information encoded in a video. The Unified Medical Language System (UMLS) developed by the National Library of Medicine (NLM) is used as a reference for medical terminology annotation. MetaMap is a highly configurable program developed to map biomedical text to the UMLS Metathesaurus and to discover Metathesaurus concepts referred to in the text. MetaMap uses a knowledge-intensive approach based on symbolic natural-language processing, and computational-linguistic techniques. The Semantic Type information is provided in the MetaMap output. It is a well-formatted semantic structure and could be used as a baseline for medical terms classification (Bodenreider et al. 2004). We create a dictionary with a list of semantic types, and a list of terms under each semantic type and examine if they capture a high percentage of medical terms. Then we make a decision whether a semantic type should be included or excluded in our annotation. We set a target precision of 80% and select a random sample of words and check how important those words are and manually eliminate words that are noisy. The selected semantic types are as follows.

Table 1. Selected Semantic Types from UMLS

| | | |
|---|---|---|
| topp (Therapeutic or Preventive Procedure) | inpo (Injury or Poisoning) | enzy (Enzyme) |
| dsyn (Disease or Syndrome) | mobd (mental or Behavioral Dysfunction) | bdsy (Body System) |
| phsu (Pharmacologic Substance) | aapp (Amino Acid, Peptide, or Protein) | antb (Antibiotic) |
| bpoc (Body Part, Organ, or Organ Component) | hcro (Health Care Related Organization) | horm (hormone) |
| neop (Neoplastic Process) | bodm (Biomedical or Dental Material) | vita (Vitamin) |
| orch (Organic Chemical) | elii (Element, Ion, or Isotope) | clnd (Clinical Drug) |
| diap (Diagnostic Procedure) | nnon (Nucleic Acid, Nucleoside, or Nucleotide) | chem (Chemical) |
| hlca (Health Care Activity) | hops (Hazardous or Poisonous Substance) | medd (Medical Device) |
| chvf (Chemical Viewed Functionally) | cgab (Congenital Abnormality) | resa (Research Activity) |
| prog (Professional or Occupational Group) | lbtr (Laboratory or Test Result) : include numbers | sosy (Sign or Symptom) |
| chvs (Chemical Viewed Structurally) | bacs (Biologically Active Substance) | inch (Inorganic Chemical) |
| lbpr (Laboratory Procedure) | drdd (Drug Delivery Device) | patf (Pathologic Function) |
| blor (Body Location or Region) | acab (Acquired Abnormality) | |

*3.3.2 Understandability*
*Is the video understandable?* PEMAT assesses video understandability based on content completeness, word choice and style, use of numbers, organization of the material, layout and design and the use of visual aids (Shoemaker et al. 2013). By following the steps for calculating video understandability score and setting a threshold of 50%, the research associates rate the 300 videos for understandability using PEMAT. The higher the score, the more understandable the material is, such as a video with an understandability score of 90% vs. one with a score of 60%.

*3.3.3 Recommendation*
*Would you recommend the video to patients for educational purposes?* We also annotate the videos based on whether the annotator would recommend the video to patients or not. When annotating videos as recommended or not recommended, the research associates do not consider other criteria such as the medical information level or understandability score.

# 4 Methods

We first generate colonoscopy-related keywords from several sources. We collect video metadata using YouTube data API. We analyze videos using Google Cloud video intelligence API. We annotate the video on medical information, video understandability, and overall recommendation. We train a bidirectional LSTM model to extract medically meaningful terms from video descriptions and narratives. Video understandability is assessed using the PEMAT framework. Finally, we build a logistic regression-based classifier to select target videos for patients' education using the two major criteria and evaluate the outcome (Liu et al. 2017).

## 4.1 Medical Entity Recognition

We leverage a deep learning-based method, specifically, the Bidirectional Long Short-Term Memory Recurrent Neural Network, to extract medical terms from the video description and assess the level of medical information. Recurrent neural networks (RNNs) are a family of neural networks that operate on sequential data such as text and speech. In the context of medical entity extraction, RNNs take as input text a sequence of vectors ($x_1, x_2,..., x_n$) and return another output sequence ($h_1, h_2,..., h_n$). The output sequence represents entity labels of the sequence at every step in the input. RNN models with word embedding input have achieved superior performance in many applications of natural language processing and understanding, such as parts-of- speech tagging, named entity recognition, and machine translation. Although RNNs can learn long sequences (long input sentences) in theory, they fail to do so in practice and tend to be biased towards their most recent inputs in the sequence. Long Short-term Memory Networks (LSTMs) tackle this issue by incorporating a memory-cell and have been shown to capture long-range dependencies. They do so with several gates that control the proportion of the input to give to the memory cell, and the proportion from the previous state to forget. LSTM RNNs have demonstrated superior performance for named entity recognition on noisy user-generated text (LeCun et al. 2015). We design an automatic cleaning process for the medical term extraction model input. For the UMLS output, we remove punctuations and symbols from the terms and split terms into single words. Then we remove stopwords from both SpaCy, NLTK and a list of most common words in English, and stipulate that the word length is greater than three.

We evaluate the popular CRF approach in information extraction as a baseline method to compare our proposed medical information extraction approach. CRF is an undirected statistical graphical model that is widely adopted for information extraction (ref). We utilize CRFsuite to train a CRF model to extract medical terms given new sentences. With the UMLS metamap preprocessing step, 1917 unique medical entities were extracted and mapped back to the video descriptions and we finally collected 19300 sentences as training data. The 285 annotated videos are divided into two parts: 80% for training and the remaining 20% for testing. Table 2 reports the performance evaluation of medical information extraction using precision (P), recall (R), and F-measure (F).

Table 2. Performance of Medical Information Extraction

|  | Precision | Recall | F-measure |
|---|---|---|---|
| **CRF** | 0.923 | 0.932 | 0.927 |
| **BLSTM RNN** | 0.959 | 0.962 | 0.960 |

The results demonstrate that the BLSTM approach for medical information extraction outperforms the baseline method. The improvement is due to the vector representation of words within their context made possible through word embedding representation of medical terminologies expressed in consumer health vocabulary.

### 4.2 Video-level Feature Extraction

Several functions of the Google Cloud Video Intelligence API are applied to detect important features of each video from the perspective of understandability.

The Speech-to-Text API transcribes the spoken word in audio format in a video or video segment into text and returns blocks of text for each portion of transcribed audio. We get the 'confidence' value from the analysis, which is an estimate between 0.0 and 1.0. It is calculated by aggregating the "likelihood" values assigned to each word in the audio. A higher number indicates a greater estimated likelihood that the individual words were recognized correctly. Those features indicate whether video materials use common, everyday language and active voice.

Analyzing videos for The Video Intelligence API can identify entities shown in video footage using the LABEL_DETECTION feature and annotate these entities with labels (tags). This feature identifies objects, locations, activities, animal species, products, and more. The confidence value is also provided within a range of [0,1]. The analysis can be compartmentalized at the segment, shot or frame level. We also apply the text detection function to capture characters that appear in a video or video segments. The API performs Optical Character Recognition (OCR) to detect visible text from frames in a video, or video segments, and returns the detected text along with information about the frame-level location and timestamp in the video for that text. Those features can imply whether videos have well-defined layout and design, proper use of visual aids and breaks or "chunks" information.

## 5 Analysis and results

### 5.1 Descriptive Summary

We fetch the top 50 videos for each search term and store the ranking of returned videos and their metadata in a database for further analysis. The videos are contributed by both individual users and reputable healthcare organizations such as Mayo Clinic, American Cancer Society, and American Nutrition Association.

Table 3. Descriptive Summary of Video Information

|  |  | mean | std | min | 25% | 50% | 75% | max |
|---|---|---|---|---|---|---|---|---|

| | | | | | | | |
|---|---|---|---|---|---|---|---|
| **Annotation** | **Medical information rating** | 0.75 | 0.44 | 0.00 | 0.00 | 1.00 | 1.00 | 1.00 |
| | **Understandability score** | 0.41 | 0.49 | 0.00 | 0.00 | 0.00 | 1.00 | 1.00 |
| | **Recommendation** | 0.38 | 0.49 | 0.00 | 0.00 | 0.00 | 1.00 | 1.00 |
| **Video level features** | **OCR confidence** | 0.87 | 0.17 | 0.00 | 0.86 | 0.89 | 0.95 | 1.00 |
| | **Number of active verbs** | 145.45 | 215.11 | 0.00 | 30.00 | 71.00 | 157.00 | 1,483.00 |
| | **Readability** | 9.44 | 5.95 | -6.80 | 6.41 | 9.42 | 12.84 | 44.18 |
| | **Number of sentences** | 41.00 | 58.79 | 0.00 | 9.00 | 21.00 | 48.00 | 399.00 |
| | **Number of shots** | 5.21 | 4.16 | 0.00 | 2.00 | 4.00 | 7.00 | 21.00 |
| | **Shot change confidence** | 0.49 | 0.15 | 0.00 | 0.44 | 0.52 | 0.59 | 0.79 |
| | **Number of summary words** | 0.20 | 0.53 | 0.00 | 0.00 | 0.00 | 0.00 | 4.00 |
| | **Transcription confidence** | 0.67 | 0.26 | 0.00 | 0.64 | 0.78 | 0.84 | 0.91 |
| | **Number of transition words** | 11.71 | 8.40 | 0.00 | 6.00 | 10.00 | 16.00 | 38.00 |
| | **Number of words** | 839.90 | 1,178.62 | 0.00 | 198.00 | 457.00 | 867.00 | 7,041.00 |
| | **Number of unique words** | 276.54 | 249.16 | 0.00 | 119.00 | 214.00 | 349.00 | 1,391.00 |
| **Metadata features** | **If the video has a title** | 1.00 | 0.00 | 1.00 | 1.00 | 1.00 | 1.00 | 1.00 |
| | **If the video has a description** | 0.95 | 0.22 | 0.00 | 1.00 | 1.00 | 1.00 | 1.00 |
| | **If the video has tags** | 0.00 | 0.00 | 0.00 | 0.00 | 0.00 | 0.00 | 0.00 |
| | **Readability** | 11.97 | 5.86 | -5.87 | 8.88 | 12.12 | 14.56 | 38.01 |
| | **Number of sentences** | 5.20 | 7.56 | 0.00 | 2.00 | 3.00 | 5.00 | 52.00 |
| | **Number of words** | 92.68 | 132.40 | 0.00 | 25.00 | 51.00 | 90.00 | 793.00 |
| | **Number of unique words** | 61.48 | 66.70 | 0.00 | 24.00 | 42.00 | 67.00 | 347.00 |
| | **Number of transition words** | 2.84 | 2.62 | 0.00 | 1.00 | 2.00 | 4.00 | 14.00 |
| | **Number of summary** | 0.02 | 0.13 | 0.00 | 0.00 | 0.00 | 0.00 | 1.00 |

|         | words                           |        |        |       |       |       |       |         |
|---------|---------------------------------|--------|--------|-------|-------|-------|-------|---------|
|         | **Number of active verbs**      | 11.58  | 18.69  | 0.00  | 2.00  | 6.00  | 11.00 | 118.00  |
|         | **Video duration**              | 377.14 | 489.04 | 11.00 | 115.00| 208.00| 390.00| 3,807.00|
| **Bi-LSTM** | **Number of unique medical terms** | 9.62   | 16.55  | 0.00  | 2.00  | 5.00  | 9.00  | 182.00  |

## 5.2 Video Classification

We train 3 different logistic regression-based classifiers to identify videos. For the first one, we use content related to video level features to classify videos into high or low medical information videos. The second classifier focuses on grouping videos into understandable ones or not. We exclude the medical information features then use the remaining features related to video understandability to train the model. We train a final logistic regression model with all features to classify videos into recommended ones or not. The feature sets provided to the classifiers are shown in the table.

**Table 4. Feature Summary of Classifiers**

| **Video recommendation classifier** | medical information rating, understandability, OCR confidence, number of active verbs(video level), readability(video level), number of sentences(video level), number of shots, shot change confidence, number of summary words(video level), transcription confidence, number of transition words(video level), number of words(video level), number of unique words(video level), if the video has a title, if the video has a description, if the video has tags, number of unique medical terms, readability(metadata ), number of sentences(metadata ), number of words(metadata ), number of unique words(metadata ), number of transition words(metadata ), number of  summary words(metadata ), number of active verbs(metadata ), video duration |
|---|---|
| **Medical information classifier** | if the video has a title, if the video has a description, if the video has tags, number of unique medical terms, readability(metadata ), number of sentences(metadata ), number of words(metadata ), number of unique words(metadata ), number of transition words(metadata ), number of summary words(metadata ), number of active verbs(metadata ), video duration, number of transition words(video level), number of words(video level), number of unique words(video level), number of active verbs(video level), readability(video level), number of sentences(video level) |
| **Video understandability classifier** | OCR confidence, number of active verbs(video level) , readability(video level), number of sentences(video level), number of shots, shot change confidence, number of summary words(video level), transcript confidence, number of transition words(video level), number of words(video level), number of unique words(video level) |

The medical information classifier has higher recall but lower precision because of the unbalanced dataset. There are 75% videos annotated as encoding a high level of medical information. The video understandability score focuses on the video organization, layout, design and use of visual aids. The dataset is reasonably balanced with about 40% videos understandable. The significant coefficients for two classifiers overlap on important features from video metadata including word count and sentence count. There are also uniquely significant features, for example, transition word count and active verb count from metadata for medical information classifier and video duration, summary word count from metadata plus unique word count, transition word count from video level data.

Table 5. Medical Information & Video Understandability Classification Evaluation Results

|  | Precision | Recall | F-measure | Overall accuracy |
|---|---|---|---|---|
| **Medical Information Classifier** | 0.709 | 1.00 | 0.830 | 0.719 |
| **Video Understandability Classifier** | 0.667 | 0.400 | 0.500 | 0.650 |

We are interested in the marginal value of combining all features together into a classifier, so we focus on the relationship between encoded medical information in videos as well as video understandability criteria and the video recommendation. Table 4 shows a summary of the logistic regression classifier for video recommendation classification. The model is trained on 228 videos and evaluated on 57 videos. The number of unique medical terms is a significant predictor for recommended videos.

Table 6. Logistic Regression Model Summary

|  | Recommendation | | Medical Information | | Understandability | |
|---|---|---|---|---|---|---|
|  | Estimate | P-value | Estimate | P-value | Estimate | P-value |
| (intercept) | -3.66 | <0.01 | -0.51 | <0.05 | -1.85 | <0.05 |
| OCR_confidence | 3.09 | <0.01 | - | - | -0.61 | 0.084 |
| understandability | 1.78 | <0.01 | - | - | - | - |
| Number of words in transcription | 1.10 | <0.05 | 0.29 | <0.05 | 1.64 | <0.01 |
| Number of sentences(metadata ) | 0.53 | <0.05 | 0.16 | 0.550 | - | - |
| Number of active verbs(video level) | 0.28 | <0.05 | 0.19 | 0.166 | 1.27 | <0.05 |
| Number of unique medical terms from Bi-LSTM model | 0.12 | <0.05 | 0.00 | 0.755 | - | - |
| Number of words(video level) | 0.00 | 0.706 | 1.26 | <0.01 | 0.50 | 0.883 |
| If the video has a description | 0.00 | 0.536 | 0.00 | 0.761 | - | - |
| If the video has tags | -0.05 | 0.093 | 0.19 | 0.420 | - | - |

| Number of sentences(video level) | -0.06 | <0.05 | 2.91 | <0.01 | 0.49 | 0.067 |
|---|---|---|---|---|---|---|
| Number of summary words(metadata ) | -0.15 | 0.128 | -0.04 | 0.078 | - | - |
| Readability(metadata ) | -0.16 | 0.074 | 0.15 | <0.05 | - | - |
| Number of transition words(metadata ) | -0.31 | 0.068 | 0.68 | 0.247 | - | - |
| Number of words(metadata ) | -0.43 | <0.05 | 0.00 | 0.204 | - | - |
| Number of unique words(metadata ) | -0.43 | <0.05 | 0.53 | 0.154 | - | - |
| Number of active verbs (metadata ) | -0.44 | <0.05 | -0.08 | 0.182 | - | - |
| Shot change confidence | -0.45 | 0.220 | - | - | -0.61 | <0.05 |
| If the video has a title | -0.45 | <0.05 | - | - | - | - |
| Number of shots | -0.46 | 0.059 | - | - | -0.87 | <0.05 |
| Number of summary words (video level) | -0.51 | <0.05 | 0.35 | 0.076 | -1.13 | <0.05 |
| Readability (video level) | -0.54 | 0.445 | -0.18 | <0.01 | -0.66 | 0.086 |
| Video duration | -0.68 | <0.05 | 0.76 | <0.05 | - | - |
| Number of unique words (video level) | -0.81 | 0.326 | 0.12 | 0.189 | -0.76 | 0.078 |
| Transcription confidence | -0.88 | <0.05 | - | - | -0.97 | <0.05 |

The performance of the classification is reported in Table 5 below. We deploy this classification model to classify the remaining videos in our collection. In the test video set, we have 24 videos classified as high medical information and 33 videos classified as low medical information.

Table 7. Video Recommendation Classification Evaluation Results

|  | **Precision** | **recall** | **F-measure** |
|---|---|---|---|
| Videos recommended | 0.917 | 0.957 | 0.936 |
| Videos not recommended | 0.970 | 0.941 | 0.955 |
|  | **Overall accuracy**: 0.947 |||

We want to identify videos that contain valid medical information for patients that consumers can also easily understand for educational purposes. There are videos that are provided by professional organizations or presented as lecture recordings by some universities containing quite a lot of medical information but are too incomprehensible for patients. Those are created for educating medical students, although they contain a large amount of medical information, they are quite challenging for patients to understand. Some videos are well-organized and formatted but are advertisements from some clinics. Some do not have audio but have well-organized text or slides showing on the screen, while others have no speech, just videos and images, which are encoded with low-level medical information. Some videos are just colonoscopy images without any explanation on the screen, subtitles or audios. In the Google Cloud Video Intelligence API, those videos have no data in the text detection and transcription detection fields. Some are taken by patients using cell phones and just represented as funny videos, for example, the hangover reaction after a colonoscopy test. Typically, those have lower confidence scores in transcription and do not have any text detected on the screen. However, some user- generated videos may not contain as much information as videos generated by professional organizations but are useful for patients who have been diagnosed with colon cancer. For example, when patients share their stories, emotional and physical changes over the years after being diagnosed with colon cancer, other patients may find these videos very helpful for their own condition.

# 6 Discussion and future work

We plan to do causal analysis with engagement measures to check the robustness of the study and evaluate the classification results. This would allow us to analyze whether the videos we recommend actually engage patients more and be effective for patient education. We intend to build on this work by employing ThinkAloud protocols and group study protocols to see how patients and educators think about the selected videos.

In this study, we also tried some other methods for medical term annotation, determining boundary decisions for medical terms and semantic type selection which did not generally perform well. Future work may include developing heuristic algorithms to preprocess the outputs of these methods. The results depend on the input terms, especially the boundary conditions. Therefore, boundary checking and correction are needed for future exploration.

Another issue for future work is to develop a pre-trained medical information extraction model for different health conditions so that patients and healthcare providers could save time on the data collection and analysis process. We also plan to refine the content coverage problem and generate educational materials for detailed topics on specific health conditions. A limitation to be acknowledged is that the annotation was conducted by three graduate students which may have introduced potential bias and needs to be evaluated further.

# 7 Conclusion

In this study, we synthesize information retrieval, deep learning and statistical methods to identify understandable and medically relevant YouTube videos on colonoscopy. We first collect video materials using keywords, extract medical terminology with a deep learning architecture, combine medical terms and other video level features from YouTube data API and google cloud intelligence video API and classify medical information encoded in YouTube videos using a logistic regression classifier. Our findings have implications for information systems researchers, healthcare professionals, patient educators, and policymakers. Based on our findings, we can also propose normative guidelines for content creators and healthcare practitioners to produce engaging and relevant patient education materials. Our study contributes to improving chronic care management by better connecting patients and caregivers with online resources that provide patient-centered self-care and decision support with digital therapeutic interventions.

# Appendix

**Appendix A. Video Search Keywords**

| colonoscopy | colonic X-rays | colon ulcer |
| --- | --- | --- |
| colonoscopy preparation | colorectal cancer | colon cancer |
| colonoscopy insertion technique | colorectal cancer test | colon cancer test |
| colonoscopy test | bowel cancer | colon cancer screening |
| colonoscopy procedure | bowel cancer test | colonoscopy anesthetic |
| colonoscopy risks | age for colonoscopy | colonoscopy pillcam |
| colonoscopy results | colon cancer prevention | colon polyps |
| colonoscopy recovery | colonoscopy side effect | rectal bleeding |
| colonoscopy cost | colon bleeding | |

## Appendix B. Sample Videos recommended or not recommended

| Video ID | Video Title | Rating | Explanation | YouTube Link |
| --- | --- | --- | --- | --- |
| vEtZh2Zi9TU | Colorectal cancer symptoms and screening guidelines | Recommended | This video explains colorectal cancer symptoms and screening guidelines. Visuals help explain how colorectal cancer can develop from a polyp inside the colon. | https://www.youtube.com/watch?v=vEtZh2Zi9TU |
| rkmHIG4EvHU | Colon cancer survivor - Beth Phillips | Recommended | In this video, a woman is diagnosed with stage IV colon cancer at age 47 – three years before colonoscopy screenings are recommended. She talks about her whole story including physical and emotional changes which gives patients an overall understanding of the disease. | https://www.youtube.com/watch?v=rkmHIG4EvHU |
| MdiI6-_kj-w | Blood in your poop: what it looks like & what it could mean | Recommended | This video introduces how poop is a valuable warning system of a problem inside people's bodies using cartoon effect. It also teaches patients the signal they should pay attention to before they flush. | https://www.youtube.com/watch?v=MdiI6-_kj-w |
| 53PawJcxrkg | Colonoscopy Results Are In 7/6/18 | Not recommended | This video is a blog taken by a patient using a smartphone who had a colonoscopy done. Although she mentions the test procedure, the whole video contains quite little medical information. | https://www.youtube.com/watch?v=53PawJcxrkg |
| SExZiM3DQD | 031 | Not | This video is a recording of a | https://www.youtube |

| w | Colonoscopy insertion technique | recommended | colonoscopy insertion procedure without any audio explanation. For patients who don't have a medical background, it's hard to acquire any useful medical information. | .com/watch?v=SExZiM3DQDw |
| QSqg9MXcYs4 | Low cost endoscopy and colonoscopy $1000 all inclusive | Not recommended | This video is an advertisement of a clinic. Even though the title contains one of the questions that patients most care about, the video doesn't contain valid medical information and should be used for education patients. | https://www.youtube.com/watch?v=QSqg9MXcYs4 |